\documentclass[aps,prd,twocolumn,showpacs]{revtex4}
\usepackage{graphicx}
\usepackage{latexsym}
\usepackage{amsmath}
\usepackage{amsfonts}
\usepackage{amssymb}
\newcommand{\be}{\begin{equation}}
\newcommand{\ee}{\end{equation}}
\newcommand{\ben}{\begin{eqnarray}}
\newcommand{\een}{\end{eqnarray}}
\begin{document}

\title{Supergravity brane worlds and tachyon potentials}
\author{D. Bazeia$^{a}$, F.A. Brito$^{b}$ and J.R. Nascimento$^{a}$}
\affiliation{$^a$Departamento de F\'\i sica, Universidade Federal da Para\'\i
ba,
Caixa Postal 5008, 58051-970 Jo\~ao Pessoa, Para\'\i ba,
Brazil
\\
$^b$Departamento de F\'\i sica,
Universidade Regional do Cariri, 63040-000 Juazeiro do Norte,
Cear\'a, Brazil}
\date{\today}

\begin{abstract}
We study massless and massive graviton modes that bind on thick
branes which are supergravity domain walls solutions in
$D$-dimensional supergravity theories where only the supergravity
multiplet and the scalar supermultiplet are turned on. The domain
walls are bulk solutions provided by tachyon potentials. Such
domain walls are regarded as BPS branes of one lower dimension
that are formed due to tachyon potentials on a non-BPS D-brane.
\end{abstract}

\pacs{11.10.Kk, 04.50.+h, 11.25.-w, 11.27.+d}

\maketitle

\section{Introduction}

The localization of gravity on brane worlds is a very active
matter of investigation in modern physics that was initiated in
\cite{rs2}. We study here background of $p$-branes created by
scalar fields of the bosonic sector of a $D$-dimensional
supergravity theory. We shall mainly explore the
localization of the graviton on thick $p$-branes \cite{sfetsos,gremm}
supported by these scalars. These thick $p$-branes are domain wall
solutions extended in $p\!=\!D-2$ dimensions.

As one knows, thick branes can be studied without adding any scalar field,
by smoothing out singular brane solutions \cite{csaba}. However, we are
interested in a supergravity theory with the contents relevant to
realize brane/string cosmology in arbitrary
$D$-dimensions where scalar fields can play important role. We
recall that such scalars play the role of tachyon fields
whose dynamics are truncated up to first order on the derivatives.
We choose some particular potentials in order to have connection
with modern cosmology. One of the many issues found in cosmology
is the accelerated expansion of our universe \cite{supernova}, and
this may have to do with dark energy. In view of this problem
new proposals of cosmological scenarios as, for instance,
quintessence \cite{kallosh_linde} and tachyon cosmology
\cite{gibbons,frolov,outros2} have been considered in the
literature.

In the present paper we focus mainly on the scalar contents of
tachyon cosmology in a supergravity theory. We consider here two
different tachyon potentials which we call type I and type II. The
tachyon potential of type I that one uses here is a potential in
the bulk, for which the global minima are at infinity
\cite{vilenkin1,bb99}. The tachyon potential of type II is
unbounded from below but there are BPS solutions, i.e., domain
wall solutions, with finite energy, connecting inflection points
of the potential \cite{bazeia2002}. These inflection points
defines an interval in which the potential engenders the
properties of being positive definite, with global minima at the
boundaries of the interval. The tachyon potentials of type I and
type II that we study are connected by a deformation function
\cite{bazeia2002}. They correspond to different and relevant
cosmological scenarios. For models which minima are at infinity
the tachyon matter is pressureless and may be considered a cold
dark matter candidate \cite{sen}. On the other hand, for
potentials that present minima at finite critical points the
tachyon matter has negative pressure and may be considered a
candidate for quintessence \cite{frolov}. Finally we should
mention that in spite of the fact that the tachyon potential of
type II considered here in general breaks the supersymmetry, one
sees that one half of the supersymmetry is preserved in the
solitonic sector, i.e., in the sector where fields configurations
are in the finite interval and also where the domain walls (thick
$p$-branes) can form.

The paper is organized as follows. In Sec.~ \ref{sug} we introduce
the $D$-dimensional supergravity theory. In Sec.~\ref{grav_modes} we
discuss the issue of graviton mode localization on $p$-branes. The
properties of the graviton localization for the backgrounds created
by the aforementioned potentials are discussed in Sec.~\ref{examples} and
in Sec.~\ref{correct}. We end the paper in Sec.~\ref{concu}, where we
present our final considerations.

\section{Supergravity action in arbitrary $D$-dimensions}
\label{sug}

We consider a supergravity Lagrangian for arbitrary
spacetimes in $D$-dimensions ($D\!>\!3$) invariant under
supersymmetry transformations, at least to lowest order in the
fermions, of the form \cite{bcy,ml} \ben \label{sugra}
S&=&\frac{1}{\kappa^{D-2}}\int
d^Dxe\Big[R-\kappa^{D-2}g^{MN}\partial_M\phi
\partial_N\phi\nonumber\\
&+&\bar{\psi}_{M\,i}\Gamma^{MNP}\nabla_N\psi_P^i
+\kappa^{D-2}\bar{\chi}_i\Gamma^M\nabla_M\chi^i\nonumber\\
&+&M(\phi)\kappa^{D-2}\bar{\chi}_i\chi^i-\kappa^{D-2}V(\phi)\nonumber\\
&+&\frac{1}{2}\kappa^{D}\partial_N\phi(\bar{\psi}_{M\,i}\Gamma^N\Gamma^M\chi^i
+\bar{\chi}_i\Gamma^M\Gamma^N\psi_M^i)\nonumber\\
&+&W_2\kappa^{D}(\bar{\psi}_{M\,i}\Gamma^M\chi^i-
\bar{\chi}_i\Gamma^M\psi_M^i)\nonumber\\
&-&(D-2)W\kappa^{D}\bar{\psi}_{M\,i}\Gamma^{MN}
\psi_N^i\nonumber\\
&+&\mbox{(higher order fermi terms)}\Big], \een where
$\kappa\!=1/\!M_*$ is the $D$-dimensional Planck length, $M_*$ is
the fundamental $D$-dimensional Planck scale,
$e\!=\det{e^A_M}\!=\!|\det\,g_{MN}|^{1/2}$, with a mostly plus
signature $(-++...+)$, and \ben \label{Vf}
&&\!\!\!\!\!\!\!\!\!\!\!\!V(\phi)=4(D-2)^2\left[\left(\frac{\partial
W}{\partial\phi}
\right)^2-\kappa^{D-2}\left(\frac{D-1}{D-2}\right)W^2\right]\!\!,\\
\label{w2}
&&W_2=(D-2)\frac{\partial W}{\partial\phi}, \\
\label{Mf}
&&M(\phi)=2(D-2)\frac{\partial^2W}{\partial\phi^2}-(D-2)\kappa^{D-2}W.
\een $W$ is the superpotential, the pair of graviton and gravitino
fields ($e^A_M$,$\psi_M^i$) corresponds to the supergravity
multiplet whereas the pair of scalar and spin-1/2 fermion fields
$(\phi,\chi^i)$ forms a scalar supermultiplet. We use $\Gamma_M$ to represent
Dirac matrices. We include an internal spinor index $i$ for generality.
The action (\ref{sugra}) is invariant under the following
supersymmetry transformations:
\ben \label{trans}
\delta e^A_M&=&-\bar{\epsilon}_i\Gamma^A{\psi}_M^i+ c.c.\;,\\
\delta\phi&=&\bar{\epsilon}_i\chi^i+c.c.\;,\\
\delta\psi_M^i&=&\nabla_M\epsilon^i+\kappa^{D-2}W\Gamma_M\epsilon^i,\\
\delta\chi^i&=&\left(-\frac{1}{2}\Gamma^M\partial_M\phi+W_2\right)\epsilon^i,
\een where $\epsilon^i$ is a local supersymmetry parameter and
\ben \label{nab}
\nabla_M\epsilon^i=\partial_M\epsilon^i+\frac{1}{4}\omega^{AB}_M
\Gamma_{AB}\epsilon^i, \qquad
\Gamma_{AB}=\frac{1}{2}[\Gamma_A,\Gamma_B].
\een
A similar consideration but exploring other issues concerning localization
of gravity on brane in supergravity has also been considered in
\cite{ortin,cv1,eto}.

For our purposes we employ the ``generalized" Randall-Sundrum
metric:
 \ben \label{RS} ds_D^2=e^{A(y)}\eta_{\mu\nu}dx^\mu
dx^\nu+dy^2, \een where $\mu,\nu\!=\!0,1,2,...,D-2$ are indices on
the ($D\!-\!2$)-brane. For this metric the Killing spinor
equations $\delta\psi^i_M=0$ and $\delta\chi^i=0$ provide us the
following equations: \ben \label{ks1}
\partial_yA&=&\mp2\kappa^{D-2}W,\\
\label{ks2}
\partial_y\phi&=&\pm2(D-2)\frac{\partial W}{\partial\phi}.
\een
Notice that we are assuming the scalar field $\phi$ and the spinor field
$\epsilon^i$ only depend on the transverse coordinate $y$. These
equations describe BPS solutions since they
come from the fermion supersymmetry transformations. Such
solutions preserve only half of the supersymmetries.

\section{Graviton modes on ($D-2$)-branes}
\label{grav_modes}

The wave function of graviton modes due to linearized gravity
equation of motion in an arbitrary number of dimensions
($D\!>\!3$) is given by \cite{justin,csaba,dan} \ben
\label{dalemb}
\partial_M(\sqrt{-g}g^{MN}\partial_N\Phi)=0,
\een where $\Phi$ describe the wave function of the graviton on
non-compact coordinates. $M,N\!=\!0,1,2,...,D-1$. This can be
found by taking into account only the traceless transverse (TT)
sector of the linear perturbation, in which the gravity equations
of motion do not couple to matter fields.

Let us consider $\Phi\!=\! h(y)M(x^\mu)$ into (\ref{dalemb}) and
the fact that ${\square}_{D-1}M\!=\!m^2M$, where $\square_{D-1}M$ is the
flat Laplacian on the tangent frame. Thus the wave equation for
the graviton through the transverse coordinate $y$ reads
\ben
\label{dalembz}
\frac{\partial_y(\sqrt{-g}g^{yy}\partial_y h(y))}{\sqrt{-g}}=
-m^2|g^{00}| h(y).
\een
This is our starting point to investigate both zero and
massive graviton modes on the branes.

Using the components of the metric (\ref{RS}) into the equation
(\ref{dalembz}) we have \ben\label{dalembz2}
\frac{1}{2}(D\!-\!1)\partial_yA\partial_y h(y)+\partial_y^2
h(y)=-m^2e^{-A(y)} h(y), \een which can be written as a
Schr\"odinger like equation by changing the metric (\ref{RS}) to a
conformally flat metric as \ben\label{RSconf}
ds_D^2=e^{A(z)}(\eta_{\mu\nu}dx^\mu dx^\nu+dz^2), \een and also by
considering the following changes of variables:
$h(y)\!=\!\psi(z)e^{-A(z)(D-2)/4}$ and
$z(y)\!=\!\int{e^{-A(y)/2}dy}$. In this way we can write
(\ref{dalembz2}) as a Schr\"odinger like equation \ben\label{sch}
-\partial_z^2\psi(z)+{\cal V}(z)\psi(z)=m^2\psi(z) \een where the
potential ${\cal V}(z)$ is given in the general form
\ben\label{Vconf} {\cal
V}(z)=\frac{(D-2)^2}{16}\,(\partial_zA)^2+\frac{D-2}{4}\,\partial_z^2A.
\een Here we notice that the potential depends only on the
geometrical variable $A(z)$ up to constant factors. It is easy to
check that for $D\!=\!5$ and $A(y)\!=\!-2k|y|$ one reproduces the
Randall-Sundrum \cite{rs2} potential.

\section{Explicit examples of tachyon potentials}
\label{examples}

We shall look for soliton solutions provided by potentials of
tachyons that live on unstable D-branes in superstring and bosonic
string theory \cite{sen,berkovits,asad} --- for D-brane/anti-D-brane systems,
the tachyon field should be complex, and this will not be considered here.
The tachyon dynamics on the
world-volume of a $p$-brane is described by the action \cite{moha,berg}
\ben
S=-\int d^{p+1}x V(T)\sqrt{1+\eta^{\mu\nu}\partial_\mu
T\partial_\nu T},
\een
where $V(T)$ is the tachyon potential, which should have the
general property of being a positive definite function with stable minima
at $|T|\to\pm\infty$. Expanding this action up to quadratic
first derivative terms gives \cite{zwie2}
\ben
S&=&-\int d^{p+1}x
V(T)\Big(1+\frac{1}{2}\eta^{\mu\nu}\partial_\mu
T\partial_\nu T+...\Big)\nonumber\\
&=&\int d^{p+1}x\Big(-\frac{1}{2}\eta^{\mu\nu}\partial_\mu
\phi\partial_\nu \phi-V(\phi)+...\Big),
\een
where in the second equation above we assumed that $V(T)\!=\!\phi_T^2$
--- here $\phi_T$ stands for $\partial_T\phi$ --- such that we can replace
$\phi_T\partial_\mu T$ to $\partial_\mu\phi$. The tachyon dynamics
is now described by the action for the scalar field $\phi$. This
action can be viewed as the scalar sector of the action
(\ref{sugra}). The scalar field here is related to the former
tachyon field $T$ and transforms as $\phi\!=\!f(T)$.

We now focus on two special examples that allow to
form domain walls solutions.

The model type I: $\phi\!=\!(1/\lambda)\, {\rm arcsinh}\,{(\lambda T)}$.
This produces a tachyon potential
\ben\label{tc_f1}
V(T(\phi))&=&\frac{1}{1+\lambda^2T^2}\nonumber\\
&=&{\rm sech}^2\,{(\lambda\phi)}.
\een
Notice that the potential above
has the property of having global minima at $T$(or $\phi)\!=\!\pm
\infty$. Potentials of this type are related to the massless limit
of supersymmetric QCD. In this limit, supersymmetric QCD coupled
to supergravity might play the role of the hidden sector of
supergravity theories \cite{seiberg}. These potentials have been
employed in models of cosmology \cite{steinhardt}.

The model type II: $\phi\!=\!T/\sqrt{1+T^2}$. This produces a
tachyon potential of the form
\ben\label{tc_f2}
V(T(\phi))&=&\frac{1}{(1+T^2)^3}\nonumber\\
&=&(1-\phi^2)^3.
\een
We see here a different asymptotic behavior, since the above potential
is unbounded below. However, the two inflection
points at $\phi\!=\!\pm1$ identify the interval $-1\leq\phi\leq1$, where
BPS solution exists \cite{bazeia2002}. Thus, if one restricts the field
to the interval $\phi\in[-1,1]$, the potential gets the required profile.

The above models are relevant to cosmology, where each one plays
a different role \cite{frolov}. For the purposes of this paper, we
are mainly concerned with the fact that these potentials have
``minima" at finite or infinite critical points. They allow for
domain walls solutions that can be regarded as BPS branes of one
lower dimension --- see \cite{berkovits} and references therein.

Now we shall look for domain walls solutions and their
gravitational localization properties. As a first example we shall consider the
model type I. We introduce a superpotential with which we are able
to find soliton solutions that connect the superpotential critical
points at infinity. Let us consider the general superpotential
\ben \label{W1} W(\phi)=
\frac{\mu}{\lambda^2}\arctan[\sinh(\lambda\phi)].
\een
By using (\ref{Vf})
we can see that (\ref{W1}) produces the potential
\ben \label{Vphi}
V(\phi)&=&4(D-2)^2\Big\{\frac{\mu^2}{\lambda^2}{\rm\
sech}^2{(\lambda\phi)}\nonumber\\
&&\!\!\!\!\!-\kappa^{D-2}\left(\frac{D-1}{D-2}\right)\frac{\mu^2}{\lambda^4}
\arctan^2[\sinh(\lambda\phi)]\Big\}.
\een
The first term in (\ref{Vphi}) comes from the derivative
\ben\label{Wf1}\frac{\partial
W}{\partial\phi}\!=\!\frac{\mu}{\lambda}{\rm\
sech}{(\lambda\phi)}
\een
and should appear even in global supersymmetry \cite{bb99}, i.e.,
when $\kappa\!=\!0$. The potential
(\ref{Vphi}) (for $\kappa\!=\!0$) up to constant factors is of the
type given in (\ref{tc_f1}).

We can deal with the system of equations (\ref{ks1}) and
(\ref{ks2}) easily because they are decoupled. The soliton
solution can be easily found by using (\ref{ks2}):
\ben
\label{sol1} \phi(\xi)=\pm\frac{1}{\lambda}{\rm arcsinh}{\,\xi},
\een
where we have considered $\xi\!\equiv\!2(D-2)\mu y$. These
solutions asymptote to infinity. Now we apply the positive solution to the
equation that describes the geometry (\ref{ks1}) --- we have set
here $\kappa\!=\!1$ for simplicity. By using the superpotential
(\ref{W1}) we find
\ben \label{A1}
A(\xi)=2[\
\ln{(1+\xi^2)^{1/2}}-\xi\arctan{\xi}\ ].
\een
We set $2\lambda^2(D-2)\!=1\!$. It is not difficulty to see that the warp
factor $e^{A(\xi)}$ falls off to zero as $|\xi|$ goes to infinity.

We now search for the conformal ``$z$-coordinate" in order
for to write down the potential (\ref{Vconf}) in the Schr\"odinger
like equation (\ref{sch}). We need to perform the integral
$z(y)\!=\!\int{e^{-A(y)/2}dy}$, i.e.,
\ben\label{z-coords}
z(y)=\frac{1}{2(D-2)\mu}\int{d\xi
e^{[\xi\arctan{\xi}-\ln{(1+\xi^2)}^{1/2}]}}.
\een
For very large $\mu$, the  {\it thin} wall limit, the integral in
(\ref{z-coords}) turns out to have approximately the simple form
$\int{\exp{(|\xi|)}d\xi}\!=\!{\rm sgn{(\xi)}}\exp{(|\xi|)}$. By
continuity in the parameter $\mu$, this means in the {\it thick}
wall regime we can approximate such integration by the function
$\sinh{\xi}$, by smoothing out back the functions ${\rm
sgn{(\xi)}}\!\to\!\tanh{\xi}$ and $|\xi|\!\to\!\ln{\cosh{\xi}}$.

All we have assumed above suffices to guarantee
integrability and inversion in (\ref{z-coords}). In this
sense, up to a constant we have set to zero, we have the
following:
\ben\label{z_coords2}
z(y)=\frac{\sinh{[\alpha\mu y}]}{\alpha\mu} \;\;\;\; {\rm or}\;\;\;\;
y(z)=\frac{{\rm \rm arcsinh}\,[\alpha\mu z]}{\alpha\mu},
\een
where $\alpha\!=\!2(D-2)$. The approximation (\ref{z_coords2}) can
actually be confirmed numerically, since the numerical integral of
(\ref{z-coords}) has a ``sinh" type profile.

The potential (\ref{Vconf}) is now written explicitly as
\ben\label{Vconf2} &&{\cal
V}(z)=\frac{1}{16}\frac{\alpha^4\mu^2\arctan^2[{\rm
arcsinh}(\alpha\mu
z)]}{1+\alpha^2\mu^2 z^2}\nonumber\\
&&+\frac{1}{4}\frac{\alpha^4\mu^3\arctan[{\rm arcsinh}(\alpha\mu
z)]z}{(1+\alpha^2\mu^2 z^2)^{3/2}}\nonumber\\
&&-\frac{1}{4}\frac{\alpha^3\mu^2}{(1+\alpha^2\mu^2 z^2)[{\rm
arcsinh}^2(\alpha\mu z)+1]}. \een This potential has the
asymptotic behavior: ${\cal V}(z\!=\!\pm\infty)\!=\!0$. This means
that the potential provides no mass gap to separate the graviton
zero mode from KK (massive) modes. At $z\!=\!0$ the potential has
a minimum with depth ${\cal V}(0)\!=\!-(1/4)\alpha^3\mu^2$. In
fact, this is a volcano type potential, which tends to the
singular  one found in the Randall-Sundrum scenario as
$\mu\!\to\!\infty$ \cite{rs2}.

The analysis for the tachyon potential of the model type II is
completely analogous to the previous case. The superpotential is
chosen to be of the form
\ben\label{tach}
W&=&\frac{3\mu}{8}\left[\frac{2\phi}{3}\frac{(a^2-\phi^2)^{3/2}}{a^2}
+\phi(a^2-\phi^2)^{1/2}\right.\nonumber\\
&&\left.+a^2\arctan{\frac{\phi}{\sqrt{a^2-\phi^2}}}\right],
\een
which is restricted to the interval $-a\!<\!\phi<\!a$ and produces
the potential
\ben\label{ptach}
&&V(\phi)=4(D-2)^2\Big\{\frac{\mu^2(a^2-\phi^2)^3}{a^4}-
\kappa^{D-2}\left(\frac{D-1}{D-2}\right)\times\nonumber\\
&&\frac{9\mu^2}{64}\left[\frac{2\phi}{3}\frac{(a^2-\phi^2)^{3/2}}{a^2}
+\phi(a^2-\phi^2)^{1/2}\right.\nonumber\\
&&\left.\;\;\;\;\;\;\;\;\;\;+a^2\arctan{\frac{\phi}{\sqrt{a^2-\phi^2}}}
\right]^2\Big\}.
\een
If gravitational effects are not included in the theory, i.e.
$\kappa=0$, the remaining term in (\ref{ptach}), which has also
been found in global supersymmetry \cite{bazeia2002}, is of the
type in the Eq.(\ref{tc_f2}).  This term is in fact related to the
remaining term of the potential (\ref{Vphi}), for $\kappa\!=\!0$,
by applying a $function$ $of$ $deformation$
$f(\phi)\!=\!a\tanh{(\lambda\phi)}$ as has been shown in
\cite{bazeia2002}. Precisely, we mean that given the tachyon
potential of the model type II
$V_{\kappa=0}(\phi)\!\sim\!(\mu^2/a^4)(a^2-\phi^2)^3$ one can obtain
the tachyon potential of the model type I
$\widetilde{V}_{\kappa=0}(\phi)\!\sim\!(\mu^2/\lambda^2)\,{\rm
sech}^2 {(\lambda\phi)}$ by just using the transformation
\ben
\widetilde{V}_{\kappa=0}(\phi)=
\frac{V_{\kappa=0}[f(\phi)]}{[f'(\phi)]^2}.
\een

As we mentioned in the introduction, the tachyon
potential of the model type II preserves half of supersymmetries
in the solitonic sector, where there are BPS
solutions defined in the interval $-a\!<\!\phi<\!a$.

We use the equation (\ref{ks2}) and the tachyon superpotential of
the model type II (\ref{tach}) to obtain the following solution
\ben\label{tachSol}
\phi=\pm a\frac{\xi}{\sqrt{\xi^2+1}}.
\een
Recall that we are considering $\xi\!=\!2(D-2)\mu y$.
This is a topological soliton solution connecting two critical
points of the superpotential $W$ which are inflections points of
$V_{\kappa=0}(\phi)$. This solution, just as the domain wall
solution found in the previous case, is regarded here as a
``thick" $p$-brane, where $y$ represents the only transverse
coordinate in $D\!=\!p+2$ dimensions.

We use equation (\ref{ks1}), setting $\kappa\!=\!1$ for
simplicity, in order to find the warp factor on this background. We get
\ben\label{warpT}
A(\xi)=2\left[\frac{1/3}{\xi^2+1}-
\xi\arctan{\xi}+\ln{\left(4\sqrt{\frac{D-2}{3}}
\right)}\right],
\een
where we have set $a\!=\!4\sqrt{{(D-2)}/{3}}$. Notice that $e^{A(\xi)}$
falls off to zero as $|\xi|$ goes to infinity. The same arguments
used before can be applied to find that here the conformal $z$-coordinate
is also given by the equation (\ref{z_coords2}).

The potential (\ref{Vconf}) can now be written explicitly as
\ben\label{Vf_tach} {\cal V}(z)=\frac{1}{4}\frac{F_3^4
F_4^2}{F_2^2}-\left[\frac{16}{3}\frac{F_3^3}{(F_1^2+1)^3 F_2^2}
+\frac{2F_4F_3^4z}{F_2^3}\right], \een where $F_1\!=\!{\rm
arcsinh}[2(D-2)\mu z]$, $F_3\!=\!D-2$ and \ben
F_2&=&\frac{[1+4(D-2)^2\mu^2z^2]^{1/2}}{\mu},\nonumber\\
F_4&=&-2\arctan{(F_1)}-\frac{2}{3}\frac{F_1(3F_1^2+1)}{(F_1^2+1)^2}\nonumber.
\een This potential as the potential (\ref{Vconf2}) has the
asymptotic behavior: ${\cal V}(z\!=\!\pm\infty)\!=\!0$. This means
that there is no mass gap. At $z\!=\!0$ the potential has a
minimum with depth ${\cal
V}(0)=-(16/3)(D-2)^3\mu^2=-(2/3)\alpha^3\mu^2$
--- compare this with the depth in the previous case. In the
following we shall look for solutions which have either zero or
massive eigenvalues.

The graviton zero mode on the ($D-2$)-brane can be found easily
since the Schr\"odinger equation (\ref{sch}) with the general
potential (\ref{Vconf}) can be written as $H\psi\!=\!m^2\psi$,
where the Hamiltonian operator is given by $H\!=\!Q^\dagger Q$,
with $Q\!=\!-\partial_z+(1/4)(D-2)\partial_z A$ --- see
\cite{dan,gremm,csaba,bws1} for further details on this issue.
Since the operator $H$ is positive definite, there are no
normalizable modes with negative energy (mass), i.e., there is no
tachyonic graviton mode. Thus the stability of the ($D-2$)-brane is ensured
and the graviton zero mode is the lowest mode in the spectrum. The
operator $Q$ annihilates the zero mode $\psi_0(z)$, i.e.
$Q\psi_0(z)\!=\!0$. This implies that the graviton zero mode is
\ben\label{zmo}
\psi_0(z)=N_0e^{\frac{D-2}{4}A(z)},
\een
where $N_0$ is a normalization constant. Notice that the metric warp factor
$e^{A(z)}$ for both cases studied above asymptotes to zero, even
in the conformal $z$-coordinate, as $|z|\!\to\!\infty$, such that
the graviton zero mode (\ref{zmo}) is normalizable, i.e.
$\int_{-\infty}^{\infty}{|\psi_0(z)|^2dz}\!<\!\infty$. We then
conclude that the graviton zero mode binds to the ($D-2$)-brane located
at $z\!=\!0$. In addition to such a mode, the potentials
(\ref{Vconf2}) and (\ref{Vf_tach}) suggest the existence of
massive modes (KK modes). These modes can affect the Newtonian
gravitational potential between two masses at the ($D-2$)-brane
located at $z\!=\!0$. In the next section we shall look for
correction to the Newtonian potential. Due to the difficulty of
integrability in (\ref{sch}) for massive modes, the
calculations that follow involve only an asymptotic analysis.

\section{Correction to the Newtonian potential at the 3-brane}
\label{correct}

The four-dimensional gravitational potential is corrected by
massive Kaluza-Klein modes \cite{rs2}. Since the KK modes
correct the Newtonian potential, the correction should be very
small, at least for the case of phenomenological interest, i.e.
$D\!=\!5$, in order to keep effectively the well known Newtonian
potential in our four-dimensional universe, the 3-brane.

We can estimate the static Newtonian potential between two
particles of mass $m_1$ and $m_2$ on the 3-brane by considering
the exchange of the zero-mode and continuum KK mode propagators
\cite{rs2}. The general form of this potential in five dimensions
\cite{sfetsos,csaba} is given by
\ben\label{newton}
&V(r)&\sim G_N\frac{m_1
m_2}{r}\nonumber\\
&&\!\!\!\!\!\!\!\!\!\!\!\!\!\!\!\!\!+M_{*}^{-3}\frac{m_1
m_2}{r}\int_{m_0}^\infty{dm\,e^{-mr}|\psi_m(0)|^2},
\een
where $M_{*}$ is the fundamental Planck scale in $D\!=\!5$ dimensions.
Notice we are integrating on the masses of the continuum modes.
The lower limit $m_0$ in the integral is set to zero when there is
no mass gap separating such continuum modes from the zero mode.

In Eq.(\ref{newton}) we see that in order to compute the
correction of the Newtonian potential we need to specify the
$\psi_m(z)$ functions for all possible $m$. In the models we are
studying these functions cannot be found analytically in a closed
form, therefore we only consider an asymptotic analysis of the
problem. In this sense, we study the quantum mechanics problem for
the potentials (\ref{Vconf2}) and (\ref{Vf_tach}), by looking for
solutions in the asymptotic limit $|z|\!\gg1\!/\alpha\mu$. In such
a limit the potentials (\ref{Vconf2}) and (\ref{Vf_tach})
asymptote to \ben\label{Vasymp} {\cal V}(z)\sim
\frac{\alpha'(\alpha'+1)}{z^2}, \een where $\alpha'=\alpha\pi/8$.
We can now solve the quantum mechanics problem for these
potentials \cite{csaba}. The solution is given by a linear
combination of Bessel functions: \ben \label{Bessel} \psi_m(z)=a_m
z^{1/2}Y_{\alpha'+1/2}(mz)+b_m z^{1/2}J_{\alpha'+1/2}(mz). \een
After considering appropriate boundary conditions it is possible
to estimate the coefficients $a_m$ and $b_m$, such that one can
estimate the value of the wavefunction  at $z\!=\!0$. The leading
term in $m$ is given by \ben\label{psi_m}
\psi_m(0)\sim\left(\frac{m}{k}\right)^{\alpha'-1}. \een The factor
of $k$ should agree with dimensional analysis. We can here
approximate this factor to the inverse of width of the domain
walls, i.e., $k\!\sim\!\mu$, which has dimension of mass.

We can now apply this result to Eq.(\ref{newton}) to find the
correction to the Newtonian potential. Integrating over all masses
by using (\ref{psi_m}), we find the four-dimensional gravitational
potential corrected as \ben \label{newton2} V(r)=G_N\frac{m_1
m_2}{r}\left(1 + \frac{C}{(k r)^{2\alpha'-1}} \right), \een where
$C$ is a dimensionless number and $G_N\!\sim\!k/M_*^3$. Notice
that the correction is highly suppressed for $k$ around the
fundamental five dimensional Planck scale and $r$ around the size
tested with gravity \cite{rs2}. Since $\alpha'\!=\!\alpha\pi/8$,
recalling that $\alpha\!=\!2(D-2)$ we have the following
situations: For the model type I, recall we have set
$2\lambda^2(D-2)\!=\!1$, thus $\alpha'\!=\!\pi/8\lambda^2$. On the
other hand, for the model type II, recall we have set
$a\!=\!4\sqrt{{(d-2)}/{3}}$, thus $\alpha'\!=\!(3/64)\pi a^2$. At
large distances, the Newtonian potential is not modified for
$\alpha'\!>\!1/2$. This implies the limits
$0\!<\!\lambda\!<\!\sqrt{\pi}/2$ and $a\!>\!(32/3\pi)^{1/2}$ on
the parameters.

Notice that in both cases discussed above, the inverse power law
of the correction is not always like $1/r^n$, where $n$ is an
integer number. For arbitrary choice of both parameters $\lambda$ and $a$
we can have rational inverse power law. In this
situation we are led to think of ``fractal dimensions" inducing
correction to the Newtonian potential. For instance if
$\lambda\!=\!(1/6)\sqrt{3\pi}$, $\alpha'\!=\!3/2$, as in the
Randall-Sundrum case \cite{rs2}, then the correction goes as
$1/r^2$. However, for $\lambda$ slightly larger than this value, we
have $1/r^p$, where $1\!<\!p\!<\!2$. On the other hand, for
$a\!=\!4\sqrt{2/\pi}$ we get the power law $1/r^2$, whereas for
$a$ slightly larger than this value, we find $1/r^p$, with
$2\!<\!p\!<\!3$.

\section{Conclusions}
\label{concu}

In this paper we have investigated the localization of zero and
massive graviton modes on thick branes supported by tachyon fields
that live on the world-volume of non-BPS D-branes. The decay
process of such objects due to tachyons ends up creating BPS
branes of one lower dimension via tachyon fields --- see \cite{berkovits}
and references therein. If we consider that the process happens
in $D\!=\!5$ dimensions, then the tachyon dynamics in the action
(\ref{sugra}) can be regarded as the tachyon dynamics on the
world-volume of a non-BPS D4-brane of superstring theory. The
brane of one lower dimension here is a BPS 3-brane and might be
our four-dimensional universe.

In the quantum mechanics problem
for the graviton spectrum on the BPS 3-brane, we have found volcano
potentials for both models type I and type II. These potentials have
just one bound state, the zero mode, and a continuum of Kaluza-Klein
modes that correct the Newtonian potential in a highly suppressed way.
With respect to the matter field excitations on the $p$-branes, it has
already been shown \cite{ml} that there are no matter zero modes on
$p$-branes solutions in the supergravity theory that we have considered.
This is because there are pre-factors like $e^{(D-n)A(z)}W(z)$,
for $n\!=\!1,3$, in the $(D-1)$-dimensional effective action that come
from the integration of the matter zero modes. Such factors are zero
due to exponential fall-off of the warp factor $e^{A(z)}$. The investigation
of matter $massive$ modes is out of the scope of the paper. Such modes
might reveal some new interesting physics and we leave this to be
explored elsewhere.

\acknowledgments

We would like to thank CAPES, CNPq, PROCAD and PRONEX  for partial
support. FAB thanks specially Justin Vazquez-Poritz and Asad Naqvi
for useful discussions.

\end{document}